# X-Ray Silicon Drift Detector-CMOS Front-End System with High Energy Resolution at Room Temperature

Bertuccio, G. ; Ahangarianabhari, M. ; Graziani, C. ; Macera, D. ; Shi, Y. ; Gandola, M. ; Rachevski, A. ; Rashevskaya, I. ; Vacchi, A. ; Zampa, G. ; Zampa, N. ; Bellutti, P. ; Giacomini, G. ; Picciotto, A. ; Piemonte, C. ; Zorzi, N.

# X-Ray Silicon Drift Detector–CMOS Front-End System with High Energy Resolution at Room Temperature

G. Bertuccio, M. Ahangarianabhari, C. Graziani, D. Macera, Y. Shi, M. Gandola, A. Rachevski, I. Rashevskaya, A. Vacchi, G. Zampa, N. Zampa, P. Bellutti, G. Giacomini, A. Picciotto, C. Piemonte, and N. Zorzi

*Abstract*—We present a spectroscopic system constituted by a Silicon Drift Detector (SDD) coupled to a CMOS charge sensitive preamplifier, named SIRIO, specifically designed to reach ultimate low noise levels. The SDD, with an active area of 13 mm$^2$, has been manufactured by optimizing the production processes in order to reduce the anode current, successfully reaching current densities between 17 pA/cm$^2$ and 25 pA/cm$^2$ at +20 °C for drift fields ranging from 100 V/cm to 500 V/cm. The preamplifier shows minimum intrinsic noise levels of 1.27 and 1.0 electrons r.m.s. at +20 °C and −30 °C, respectively. At room temperature (+20°C) the $^{55}$Fe 5.9 keV and the pulser lines have 136 eV and 64 eV FWHM, respectively, corresponding to an equivalent noise charge of 7.4 electrons r.m.s.; the noise threshold is at 165 eV. The energy resolution, as measured on the pulser line, ranges from 82 eV FWHM (9.4 electrons r.m.s.) at +30 °C down to 29 eV FWHM (3.3 electrons r.m.s.) at −30 °C.

## I. Introduction

### A. Semiconductor Drift Detectors

SEMICONDUCTOR drift detectors (SDDs) were invented by E. Gatti and P. Rehak in 1983 [1], [2] and suddenly opened new perspectives in radiation and particle detection and spectroscopy because of the higher energy resolution achievable with respect to traditional junction diodes. Fig. 1(a) shows a cross section of a cylindrical-type SDD: the p$^+$ cathodes are biased to fully deplete the silicon bulk and to generate an electrostatic potential energy for the electrons, in Fig. 1(b), whose minimum stays at a small n$^+$ electrode (anode) usually placed at the center of the device. The electrons generated by the radiation are so driven toward the anode and collected there. The SDD's anode capacitance is extremely low ($\approx$ 0.1 pF), because of its small geometrical dimension and of the fully depleted semiconductor, and independent from the detector active area. Therefore, SDDs can be designed with large sensitive area (up to several cm$^2$) but with the output capacitance of a small pixel. Differently, a traditional semiconductor detector based on planar Schottky or pn junction has a capacitance directly proportional to the junction area with a specific capacitance of about 34 pF/cm$^2$ for a 300 $\mu$m thick detector. Since the noise of a detection system based on a junction semiconductor detector significantly decreases by lowering the capacitance of the detector output electrode, SDDs have brought a breakthrough with respect to traditional detectors for which a tradeoff between large sensitive area, low noise, and number of readout channels were strictly necessary. In addition, the very small capacitance of the SDD anode not only reduces the contribution of system series noise but also shortens the optimum shaping time $\tau_{\mathrm{opt}}$ so that the contribution of the parallel white noise, arising from the detector current, is reduced as well. Besides, SDDs allow also low noise operation at high photon/particle rate because of the shorter $\tau_{\mathrm{opt}}$.

In the last 30 years, a worldwide research and development activity has been carried out to study and design different topologies of X-ray SDDs which nowadays are widely used in many scientific experiments as well as in several commercially available X-ray spectrometers for scientific, industrial and medical applications [3], [4].

### B. The Limit of the Detector Operating Temperature

One of the limits of all the semiconductor X-ray detectors with active areas above 1 mm$^2$, including SDDs, is their relatively high dark current, and consequently high noise, at room temperature, so that cooling the device is strictly necessary in order to achieve a high energy resolution. This is valid also for wide bandgap compound semiconductors such as CdTe and CdZnTe for which room temperature operation is limited to spectroscopy of high energy photons (50 keV–10 MeV) only,

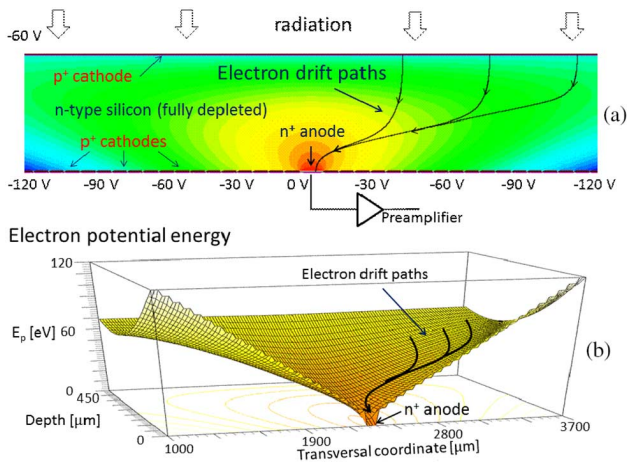

Fig. 1. (a) Section of a cylindrical Silicon Drift Detector; (b) Potential energy of the electrons in the SDD. The electrons generated by the radiation drift toward the anode electrode, where the minimum potential energy is placed by means of a proper bias of a set of cathodes.

TABLE I
X-RAY SDD PERFORMANCE AT ROOM TEMPERATURE

| Reference | year | Detector area (mm$^2$) | Temperature (°C) | 5.9 keV $^{55}$Fe FWHM (eV) | ENC (e- r.m.s.) | Peaking time (µs) |
|---|---|---|---|---|---|---|
| This work | 2015 | 13 | +20 | 136 | 7.4 | 1.4 |
| [9] | 2014 | 13 | +21 | 141 | 8.6 | 0.8 |
| [10] | 2012 | 25 | +25 | 260 | 27 | 0.1 |
| [11] | 2001 | 10 | +25 | 300 | -- | -- |
| [12] | 1996 | 3.5 | +27 | 220 | 21 | 0.5 |
| [13] | 1994 | 2 | +24 | (374) | 41 | -- |
| [14] | 1994 | 1.5 | +20 | 327 | -- | 7 |
| [14] | 1994 | 0.5 | +20 | 267 | -- | 7 |
| [15] | 1986 | 78.5 | R.T. | (940) | 110 | 0.25 |

where energy resolutions in the keV range or even higher are adequate.

The upper limit of the operating temperature of SDDs is determined by the anode current density, which currently ranges between few nA/cm$^2$ down to 200 pA/cm$^2$ at room temperature for both typical and state of the art detectors [5]. Practically, SDDs for X-ray spectroscopy operate at temperatures between $-20$ °C and $-55$ °C in order to reduce the anode current down to tens of fA, so reaching the highest energy resolution.

### C. The SDD Front-End Electronics

The Front-End Electronics (FEE) is a fundamental part of an X-ray spectroscopic system; its importance in achieving the ultimate performance from the system is as high as the detector quality. Although Junction Field Effect Transistors (JFETs) are able to guarantee high performance due to their ultra low $1/f$ noise, the very low capacitance of SDDs allows using also CMOS Charge Sensitive Preamplifiers (CSP) because the size scalability of their input transistor permits a much better capacitive matching with the detector. In 1997 Rehak et al. proved this solution with excellent results [6]. Further research have brought to CMOS CSP with few electrons noise [7] and, some year later, this has been applied to SDDs readout as an alternative to JFETs [8].

### D. Motivation and Objective of this Work

High resolution X-ray spectroscopy in the 0.1 keV–20 keV energy range with a system operating at room temperature is appealing because of the advantage of avoiding cooling systems, which are relatively bulky, heavy, with power consumption in the Watt range and require vacuum or sealed dry enclosures. SDDs are the most promising candidate to achieve this goal and have been tested at room temperature since 1986 to present. Some progress has been done in term of energy resolution and larger active area, as shown in Table I, but the energy resolution was mostly limited by the leakage current of the detector or by the noise of the FEE. Till a few years ago, the best published result at room temperature was 260 eV FWHM at $+25$ °C at 5.9 keV ($^{55}$Fe) for a 25 mm$^2$ SDD [10], which is anyhow well far away from the resolution of 123 eV–125 eV FWHM obtained with peltier-cooled SDDs [5], [16], [17]. Last year, the first achievements of the our research activity in designing SDDs and FEE with ultimate low noise was presented [9], demonstrating for the first time the capability for a SDD-CMOS FEE system to operate up to room temperature with the highest energy resolution (141 eV FWHM) possible so far, approaching the value achievable with cooled devices. In this paper, we present improved and extended experimental results of the SDD-CMOS FEE system together with their detailed analysis: in Section II the detector design, fabrication and electrical characterization are presented; Section III is devoted to the Front-End Electronics and Section IV to the system characterization; an analysis of the system's noise is presented in Section V.

## II. THE SILICON DRIFT DETECTOR

### A. Fabrication Technology

The detector was realized on a floating zone silicon wafer with (100) crystal orientation, n-type, 150 mm diameter, 450 µm thick.

The fabricated SDD is hexagonal with an active area of 13 mm$^2$ and the anode in the center (Fig. 2). A continuous p$^+$ cathode constitutes the entrance window on one side of the detector. The cathodes on the anode-side are biased through an integrated resistive voltage divider in order to create the drift field. The hexagonal structure is functional to a matrix configuration [18]. An innovative and proprietary technological process (LC-SDD/low current SDD)–driven also by the need of very large area SDDs [19]—has been developed at FBK to reduce the reverse currents of the pn junctions in order to minimize the dark current at the anode.

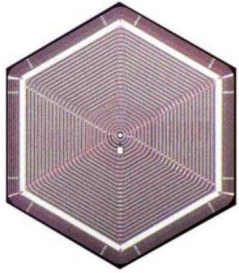

Fig. 2. Photo of anode-side of the manufactured hexagonal Silicon Drift Detector used in the spectroscopic system. The active area is 13 mm$^2$.

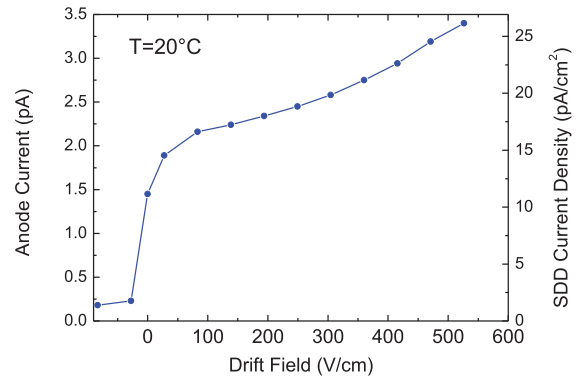

Fig. 4. Anode current of LC-SDD and current density (right y-axis) measured at +20 °C as function of the drift field.

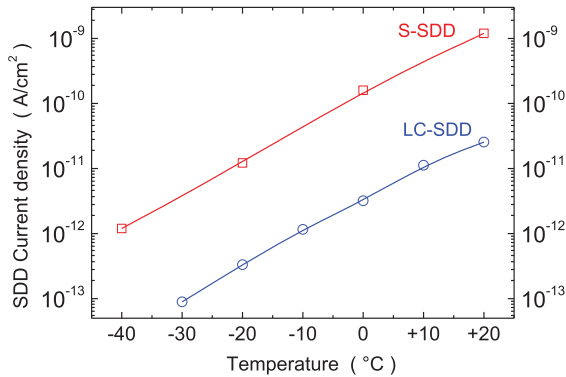

Fig. 3. Current densities measured at the anode of two Silicon Drift Detectors in operative conditions ($E_{\mathrm{DRIFT}} = 500$ V/cm) as function of the temperature. The SDD fabricated with the new low current process (LC-SDD) shows an anode current about 50 times lower than the SDD fabricated with standard process (S-SDD).

### B. Electrical Characterization

The SDD has been glued on the printed circuit board together with the charge sensitive preamplifier (described in Section III), placed in thermostatic chamber and biased at the operating conditions and firstly checked with a radioactive source. The anode current ($I_{anode}$) has been measured as a function of the temperature. Since the CSP operates in pulsed-reset mode, the anode current is periodically integrated into the feedback capacitance, so that $I_{anode}$ can be determined from the slope of the CSP output voltage. The feedback capacitance has been accurately measured from the amplitude of the signals generated by the 5.9 keV line of a $^{55}$Fe radioactive source. The results of the measurements are shown in Fig. 3 in terms of current density determined at a drift field of 500 V/cm. The same figure also shows the current density measured on a similar SDD fabricated with a standard process (S-SDD). At a temperature T = +20 °C the current density of the LC-SDD is J = 25 pA/cm$^2$ while it is J = 1.2 nA/cm$^2$ for the S-SDD showing a significant reduction by a factor 48 achieved with the LC-process. The slope of the two I-T curves are approximately the same and so almost the same reduction factor is also observed at all the other temperatures. At T = −30 °C the measured current densities are J = 4 pA/cm$^2$ and J = 90 fA/cm$^2$ for the standard and LC process, respectively.

It can be observed that the S-SDD must be cooled down to T=−14 °C in order to reach the same anode current the LC-SDD shows at T = +20 °C, so the new technology allows to gain $\Delta T = +34$ °C in terms of operating temperature to have the same detector noise. Considering that the lowest anode current density of SDDs declared so far is 200 pA/cm$^2$ at room temperature [17], this technology sets a new state of the art lowering the current density by an order of magnitude. Since the parallel noise *ENC* contribution depends on the root square of the anode current, a significant improvement by a factor 3 in this noise component is expected.

Fig. 4 shows the anode current measured at T = +20 °C as a function of the drift field, which has been changed by setting the outmost drift cathode bias voltage while keeping both the first and the entrance window cathodes bias voltages constant. An increase of the current density from 17 pA/cm$^2$ to 25 pA/cm$^2$ as the drift field is changed from 100 V/cm to 500 V/cm is measured. Such a growth can be explained with the increasing of the generation current due to the widening of the depleted intercathode surfaces and/or due to the increasing of the charge generation rate in the regions in close proximity to the voltage divider, where a local rising of the temperature is expected due to the increment of the electrical power in the integrated resistors.

Assuming that the anode current is dominated by thermal generation of charge carriers in the depleted silicon, we can write [20], [21]

$$I_{anode} = AW\frac{n_i}{2\tau_o} = \frac{AW}{2\tau_o}\sqrt{N_{Co}N_{Vo}}\left(\frac{T}{T_o}\right)^{3/2} e^{-\frac{E_g}{2kT}} \quad (1)$$

in which $A$ and $W$ are the area and the thickness of the depleted region, respectively; $n_i$ is the intrinsic carrier concentration, $\tau_o$ is the effective carrier lifetime, $N_{Co} = 2.86 \times 10^{19}$ cm$^{-3}$ and $N_{Vo} = 3.1 \times 10^{19}$ cm$^{-3}$ are the effective density of states in the conduction and valence bands, respectively, at the reference temperature $T_o = 300$ K [22]; $E_g$, $k$, and $T$ are the bandgap energy, the Boltzmann constant, and the absolute temperature. The experimental data have been fitted to (1) as shown in Fig. 5 finding an acceptable agreement, in a first approximation. The carrier effective lifetime derived from the fitting is $\tau_o = 1.16$ s, which reveals a significant improvement with respect the value of 15 ms originally measured by J. Kemmer and collaborators in 1982 with the first silicon detector based on pn junctions made using planar technology [23].

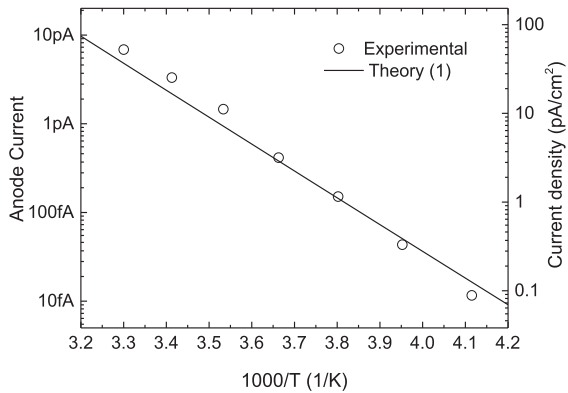

Fig. 5. Anode current vs. 1000/T. An acceptable agreement with the theoretical generation current [see (1)] is found, deriving a carrier effective lifetime of $\tau_o = 1.16$ s.

## III. THE SIRIO PREAMPLIFIER

The front-end electronics of our system is constituted by an ultra low-noise CMOS Charge Sensitive Preamplifier named SIRIO, which is the result of several years of research activity focused on achieving the lowest noise in CMOS CSP. Its first version, designed for SiC pixel detectors, demonstrated an $ENC = 3.9$ electrons r.m.s. at room temperature [7]. In this experiment we used the third generation of such CSPs (SIRIO-3G), which has an intrinsic equivalent noise charge (ENC) around 1.3 electron r.m.s. at room temperature and even lower than 1 electron r.m.s. at $-30\ °C$. SIRIO is a fully integrated CSP designed in 0.35 $\mu$m CMOS technology, it operates in pulsed-reset mode and has a power consumption of about 10 mW, a detailed description of SIRIO can be found in [24]. A SIRIO-3G chip, whose area is smaller than 0.4 mm$^2$, has been glued onto the detector in close proximity to the anode in order to minimize the bonding wire length and so the associated stray capacitance. We have tested 3 samples of this system obtaining very similar results; this paper reports the results of one of these.

## IV. EXPERIMENTAL

### A. Experimental Setup

The detector-preamplifier system was placed inside a metal box in a thermostatic chamber. The system was operating in dry air, obtained by fluxing nitrogen and by hygroscopic salts, the temperature was constantly measured by a thermocouple. A precision pulser (Tektronix AFG3022B) was used to inject calibrated charge pulses through the test capacitance ($C_{TEST}$) integrated in the CSP in order to directly measure the electronic noise of the system. The preamplifier output signal have been processed by a digital pulse processor (Amptek PX5) implementing a triangular shaping with peaking times settable from 0.4 $\mu$s to 102 $\mu$s.

### B. Readout Electronics Characterization

The readout electronics has been tested before connecting the detector anode to the preamplifier input in order to measure the noise of the system without the detector contribution. The ENC has been measured acquiring artificial spectra by injecting a sequence of two calibrated charge pulses by means of the pulser. The injected charge has been first estimated by considering the nominal value of $C_{TEST}$ and then accurately determined when the detector was connected, by measuring $C_{TEST}$ using a $^{55}$Fe spectrum. Fig. 6 shows the spectrum acquired at T = $+22\ °C$ by injecting two sets of charge pulses of about 50 and 100 electrons, equivalent to about 183 eV and 367 eV deposited in silicon. At the optimum peaking time of 51 $\mu$s a minimum FWHM of 11 eV has been measured, corresponding to ENC = 1.27 electrons r.m.s. The threshold of the noise is at only 5.6 electrons, corresponding to an energy of 21 eV. The $ENC$ was measured also at T = $-30\ °C$, resulting in a minimum noise of 1.0 electrons r.m.s. (8.7 eV FWHM). The $ENC$ of the system versus the peaking time at T = $+22\ °C$ and T = $-30\ °C$ before connecting the detector is shown in Fig. 10, indicated for simplicity as "SIRIO preamplifier" although it represents the system noise, disturbances and interference included, excluding only the contribution of the detector and of the preamplifier-detector connection.

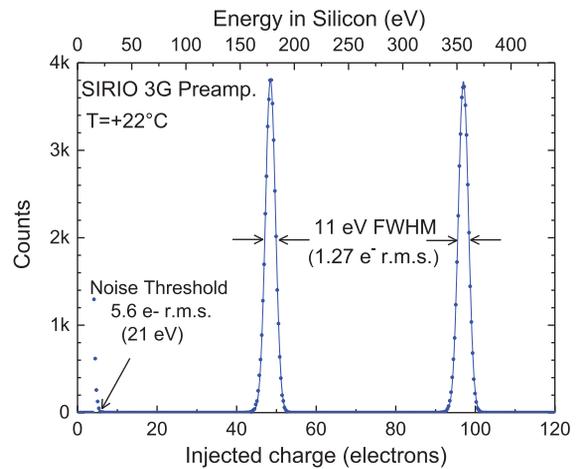

Fig. 6. Artificial spectrum acquired at $+22\ °C$ before connecting the preamplifier input to the detector anode. The measured ENC of 1.27 e$^-$ r.m.s. (11 eV FWHM) represents the system noise, disturbances and interference included, but without the detector contribution. The noise threshold is at only 5.6 electrons (21 eV).

### C. System Characterization with X-rays

After connecting the anode at the preamplifier input by wire bonding, the SDD-SIRIO system has been characterized with the $^{55}$Fe and the pulser lines. No photon collimation was used during irradiation and the total photon rate was about $2 \times 10^3$ s$^{-1}$. Fig. 7 shows the best spectrum of $^{55}$Fe acquired at room temperature (T = $+20\ °C$) with a peaking time of 1.4 $\mu$s. The width of the 5.9 keV line is 136 eV FWHM and the pulser line width is 64 eV FWHM, corresponding to 7.4 electrons r.m.s.; the noise threshold is at 165 eV (45 electrons). To our best knowledge, this is the highest energy resolution ever measured with a semiconductor detector of such area (13 mm$^2$) and operated at room temperature. Comparing this result with previously published values for 5.9 keV FWHM reported in Table I, the significant progress can be appreciated.

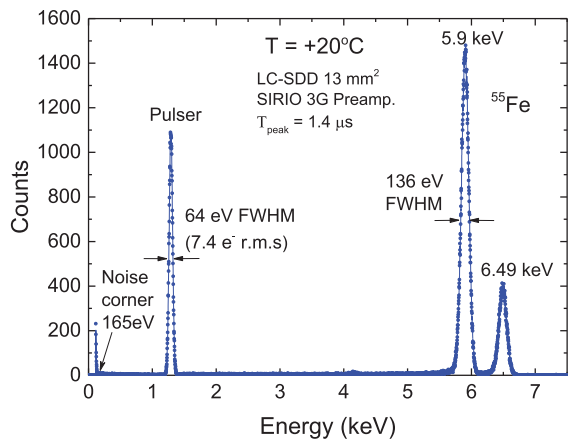

Fig. 7. Spectrum of $^{55}$Fe acquired at $+20\,°C$ and at the optimum peaking time (1.4 $\mu$s). The pulser line width is 64 eV FWHM, corresponding to 7.4 electrons r.m.s.

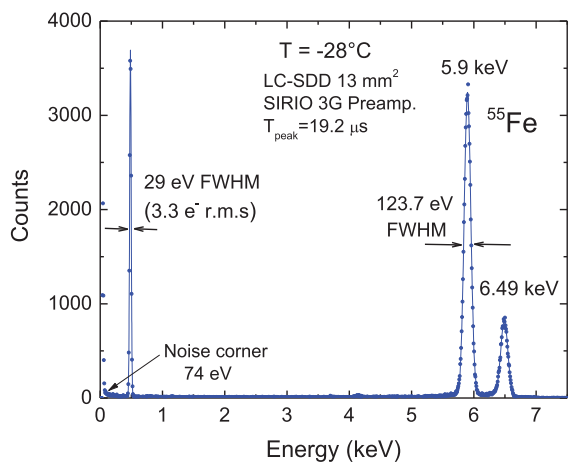

Fig. 8. Spectrum of $^{55}$Fe acquired at $-28\,°C$ and at the optimum peaking time (19.2 $\mu$s). The pulser line width is 29 eV FWHM, corresponding to 3.3 electrons r.m.s.

The system was also tested at different temperatures from T $= +30\,°C$ down to T $= -30\,°C$. Fig. 10 shows the pulser line FWHM and the corresponding *ENC* as a function of the peaking time at the different operating temperatures and Table II reports the minimum line width measured at each temperature and the corresponding peaking time. The system noise at the optimum peaking times, ranges from 82 eV FWHM at T $= +30\,°C$ down to 29 eV FWHM at T $= -30\,°C$. The $^{55}$Fe spectrum acquired at T $= -28\,°C$ is shown in Fig. 8: the width of the 5.9 keV line is 123.7 eV FWHM and the pulser line width is 29 eV FWHM, equivalent to 3.3 electrons r.m.s. at a peaking time of 19.2 $\mu$s. The peak to valley ratio in the $^{55}$Fe spectrum was not high ($\approx 10^3$) because of the absence of a photon collimator, so that the FWHM can be probably improved even further with the use of a well-designed collimator.

## V. SYSTEM NOISE ANALYSIS

### A. Noise Components Analysis

An analysis of the measured noise has been done to determine the noise contributions of the detector and of the readout electronics and to find the limiting factor to the performance of the system. The analysis has been performed by considering the classical model for the equivalent noise charge:

TABLE II
LC-SDD-SIRIO: SYSTEM ENERGY RESOLUTION AND NOISE

| T (°C) | $^{55}$Fe 5.9 keV LINE WIDTH (eV FWHM) | PULSER LINE WIDTH (eV FWHM) | ENC (e- r.m.s.) | Peaking Time ($\mu$s) |
|---|---|---|---|---|
| +30 | 148 | 82 | 9.4 | 0.8 |
| +20 | 136 | 64 | 7.4 | 1.4 |
| +10 | 133 | 53 | 6.1 | 2.4 |
| 0 | 129 | 44 | 5.0 | 4.8 |
| -10 | 129 | 41 | 4.7 | 9.6 |
| -30 | 123.7 | 29 | 3.3 | 19.2 |
| READOUT ELECTRONICS | | | | |
| +22 | -- | 11 | 1.27 | 51.2 |
| -30 | -- | 8.7 | 1.0 | 102.4 |

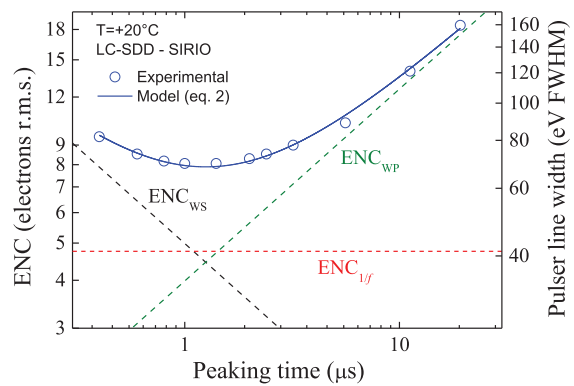

Fig. 9. LD$-$SDD$+$SIRIO system at room temperature ($+20\,°C$): Equivalent Noise Charge (left y-axis) and corresponding pulser line width (right y-axis) versus peaking time. The main noise components are derived by fitting the experimental data to (2).

$$ENC_{tot} = \sqrt{ENC_{ws}^2 + ENC_{1/f}^2 + ENC_{wp}^2} \qquad (2)$$

in which the total noise is given by the white series ($ws$), $1/f$ and white parallel ($wp$) components, whose detailed expressions can be found in [25]. Fig. 9 shows the fitting to (2) of the experimental *ENC* vs. $T_{peak}$ data taken at T $= +20\,°C$. The model well describes the experimental data allowing to precisely disentangle the three noise components. From the white series noise a total capacitance of about 130 fF at the system input can be derived, which agrees with an estimated detector capacitance of about $C_D = 60$ fF and a wire bonding capacitance ($C_W$) toward the signal ground of a few tens of fF. A leakage current of 3.8 pA is calculated from the white parallel component, in good agreement with the value measured from the slope of the preamplifier output; the $1/f$ component is contributing for about 4.8 electrons, which is almost the same value given by the series and parallel noise at the optimum peaking time of 1.2 $\mu$s.

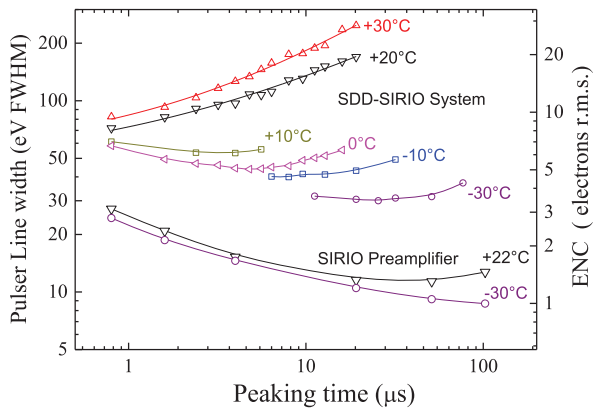

Fig. 10. Pulser line width and ENC for the whole system and of the SIRIO preamplifier alone, measured at different temperatures as function of the peaking time of a triangular pulse shaping.

*B. Discussion of Detector and Front-End Noise Contributions*

The noise of the system has been measured before and after connecting the detector anode to the preamplifier input. The results have been reported in Fig. 10: the *ENC* measured before the connection of the detector ranges from 3 e$^-$ r.m.s. (26 eV FWHM) at T = +22°C and $T_{peak}$ = 0.8 $\mu$s down to 1.0 e$^-$ r.m.s. (8.7 eV FWHM) at T = −30 °C and $T_{peak}$ = 102 $\mu$s. Such noise includes mainly the preamplifier intrinsic noise (so the curves have been indicated as SIRIO preamplifier) but also any noise component injected from the power supplies and any disturbance coming from the surrounding environment. Comparing the noise measured before and after connecting the detector, it is clear that the connection of the detector brings to a significant increase of the system noise; this is due to the relevant growth of both the parallel and series noises arising from the detector anode current and from the detector and wire bonding capacitances, respectively. This becomes even more evident if we consider that the *ENC* contributions sum quadratically: at T = −30 °C, for example, *ENC* grows from 1.0 to 3.3 electrons r.m.s., that is a variation of an order of magnitude. This increment is due to several effects. As far as the parallel noise is concerned, the preamplifier is practically noiseless at all the temperatures, since the leakage current at its input ranges from 1 fA at T = +20 °C down to few aA at T = −30 °C, so that the whole parallel noise of the system comes from the detector. The series noises (white and $1/f$) have their origin in the preamplifier (mainly from the input transistor), but in the system their intensities are strongly determined by the capacitance of the detector ($C_D$) and of the wire connection ($C_w$). In Fig. 10, it can be observed the significant increase of both the series noise components when the detector is connected, but in an ideal case of $C_D + C_w = 0$ we wouldn't observe any increase in these noise components. So, we can conclude that the detector and the preamplifier/detector wire connection, are significantly determining the system energy resolution at all temperatures. At the same time, in principle, also the preamplifier design can be further improved to reduce the *ENC* sensitivity to the input load capacitance.

## VI. CONCLUSIONS

We have designed and experimentally characterized an X-ray spectroscopic system constituted by a Silicon Drift Detector and a CMOS integrated preamplifier (SIRIO) demonstrating that very high energy resolution—with pulser linewidth equal or less than 82 eV FWHM—can be obtained even with the system operating at room temperature. This results have been made possible by two parallel and concurrent research activities: the first was related to an improved detector manufacturing technology which allowed to decrease the reverse current density of the pn junctions at ultra-low levels (anode current density lower than < 25 pA/cm$^2$ at room temperature); the second was an advanced design of a CMOS charge sensitive preamplifier showing an intrinsic noise close to 1 electron r.m.s. at room temperature.

The energy resolution measured at +20 °C (136 eV FWHM at 5.9 keV from $^{55}$Fe) is not so far from that one (124 eV FWHM) achieved with state of the art system cooled between −20 °C and −50 °C using Peltier cooling systems. Besides, the measured energy resolution is the highest ever achieved at room temperature by any semiconductor detector of comparable active area and opens new perspectives in the technological implementation and applications of these types of X-ray spectrometers.

ACKNOWLEDGMENT

This research has been supported by Politecnico di Milan, the National Institute of Nuclear Physics (INFN) and Fondazione Bruno Kessler (FBK). The authors thank S. Masci for system assembly and wire bondings, and A.C. Huber and R.H. Redus for the optimized digital pulse processor used in these experiments.

REFERENCES

[1] E. Gatti and P. Rehak, "The concept of a solid state drift chamber," in *Proc. 1983 DPF Workshop Collider Detectors: Present Capabilities and Future Possibilities*, Berkeley, CA, USA, Feb.–Mar. 28–4, 1983, p. 94720, Lawrence Berkeley Laboratory, Univ. California.
[2] E. Gatti and P. Rehak, "Semiconductor drift chamber—An application of a novel charge transport scheme," *Nucl. Instrum. Methods Phys. Res.*, vol. 225, no. 3, pp. 608–614, Sep. 1, 1984.
[3] E. Gatti and P. Rehak, "Review of semiconductor drift detectors," *Nucl. Instrum. Methods Phys. Res.*, vol. 541, pp. 47–60, 2005.
[4] C. Guazzoni, "The first 25 years of Silicon Drift Detectors: A personal view," *Nucl. Instrum. Methods Phys. Res.*, vol. 624, pp. 247–254, 2010.
[5] R. Redus, A. Huber, J. Pantazis, and T. Pantazis, "Enhanced energy range thermoelectrically cooled silicon X-ray detectors," in *IEEE Nucl. Sci. Symp. Conf. Rec.*, 2011, pp. 580–585.
[6] P. O'Connor, G. Gramegna, P. Rehak, F. Corsi, and C. Marzocca, "CMOS preamplifier with high linearity and ultra low noise for X-ray spectroscopy," *IEEE Trans. Nucl. Sci.*, vol. 44, no. 3, pp. 318–325, 1997.
[7] G. Bertuccio and S. Caccia, "Progress in ultra-low-noise ASICs for radiation detectors," *Nucl. Instrum. Methods Phys. Res.*, vol. 579, no. 1, pp. 243–246, Aug. 21, 2007.
[8] L. Bombelli, C. Fiorini, T. Frizzi, R. Alberti, and A. Longoni, "CUBE, A low-noise CMOS preamplifier as alternative to JFET front-end for high-count rate spectroscopy," in *Proc. 2011 IEEE Nucl. Sci. Symp. Med. Imag. Conf. (NSS/MIC)*, pp. 1972–1975.
[9] G. Bertuccio, M. Ahangarianabhari, C. Graziani, D. Macera, Y. Shi, A. Rachevski, I. Rashevskaya, A. Vacchi, G. Zampa, N. Zampa, P. Bellutti, G. Giacomini, A. Picciotto, and C. Piemonte, "A Silicon Drift Detector-CMOS front-end system for high resolution X-ray spectroscopy up to room temperature," *J. Instrum.*, vol. 10, Jan. 2015, Art. no. P01002.
[10] L. Bombelli, C. Fiorini, T. Frizzi, R. Alberti, and R. Quaglia, "High rate X-Ray spectroscopy with "CUBE" preamplifier coupled with Silicon Drift Detector," in *IEEE Nucl. Sci. Symp. Med. Imag. Conf. Rec. (NSS/MIC)*, 2012, pp. 418–420.

[11] P. Lechner, C. Fiorini, R. Hartmann, J. Kemmer, N. Krause, P. Leutenegger, A. Longoni, H. Soltau, D. Stotter, R. Stotter, L. Struder, and U. Weber, "Silicon drift detectors for high count rate X-ray spectroscopy at room temperature," *Nucl. Instrum. Methods Phys. Res.*, vol. 458, pp. 281–287, 2001.

[12] P. Lechner, S. Eckbauer, R. Hartmann, S. Krisch, D. Hauff, R. Richter, H. Soltau, L. Strüder, C. Fiorini, E. Gatti, A. Longoni, and M. Sampietro, "Silicon drift detectors for high resolution room temperature X-ray spectroscopy," *Nucl. Instrum. Methods Phys. Res. A*, vol. 377, no. 2–3, pp. 346–351, Aug. 1, 1996.

[13] R. Hartmann, D. Hauff, S. Krisch, P. Lechner, G. Lutz, R. H. Richter, H. Seitz, L. Struder, G. Bertuccio, L. Fasoli, C. Fiorini, E. Gatti, A. Longoni, E. Pinotti, and M. Sampietro, "Design and test at room temperature of the first Silicon Drift Detector with on chip electronics," in *Proc. IEEE Int. Electron Devices Meet.*, San Francisco, CA, USA, Dec. 11–14, 1994.

[14] P. Jalas, A. Niemela, W. Chen, P. Rehak, A. Castoldi, and A. Longoni, "New results with semiconductor drift chambers for X-ray spectroscopy," *IEEE Trans. Nucl. Sci.*, vol. 41, no. 4, pp. 1048–1053, 1994.

[15] P. Rehak, J. Walton, E. Gatti, A. Longoni, M. Sampietro, J. Kemmer, H. Dietl, P. Holl, R. Klanner, G. Lutz, and A. Wylie, *Nucl. Instr. Meth.*, vol. 248, no. 1986, pp. 367–378, 1986.

[16] R. Quaglia, L. Bombelli, C. Fiorini, M. Occhipinti, P. Busca, G. Giacomini, F. Ficorella, A. Picciotto, and C. Piemonte, "New developments of SDD-based X-ray detectors for the siddharta-2 experiment," in *Proc. IEEE Nucl. Sci. Symp. Med. Imag. Conf. (NSS/MIC)*, 2013.

[17] D. M. Schlosser, P. Lechner, G. Lutz, A. Niculae, H. Soltau, L. Struder, R. Eckhardt, K. Hermenau, G. Schaller, F. Schopper, O. Jaritschin, A. Liebel, A. Simsek, C. Fiorini, and A. Longoni, "Expanding the detection efficiency of Silicon Drift Detectors," *Nucl. Instrum. Methods Phys. Res.*, vol. 624, pp. 270–276, 2010.

[18] A. Rachevski, G. Zampa, N. Zampa, I. Rashevskaya, A. Vacchi, G. Giacomini, A. Picciotto, A. Cicuttin, M. L. Crespo, and C. Tuniz, "X-ray spectroscopic performance of a matrix of silicon drift diodes," *Nucl. Instrum. Methods Phys. Res.*, vol. 718, pp. 353–355, 2013.

[19] A. Rachevski, G. Zampa, N. Zampa, R. Campana, Y. Evangelista, G. Giacomini, A. Picciotto, P. Bellutti, M. Feroci, C. Labanti, C. Piemonte, and A. Vacchi, "Large-area linear Silicon Drift Detector design for X-ray experiments," *J. Instrum.*, vol. 9, 2014, Article no. P07014.

[20] A. S. Grove, *Physics and Technology of Semiconductor Devices*. New York: Wiley, 1967.

[21] S. M. Sze, *Physics of Semiconductor Devices*, 2nd ed. New York: Wiley, 1981, pp. 19–91.

[22] M. A. Green, "Intrinsic concentration, effective densities of states, and effective mass in silicon," *J. Appl. Phys.*, vol. 67, pp. 2944–2954, 1990.

[23] J. Kemmer, P. Burger, R. Henck, and E. Heijne, "Performance and applications of passivated ion-implanted silicon detectors," *IEEE Trans. Nucl. Sci.*, vol. 29, no. 1, pp. 733–737, 1982.

[24] G. Bertuccio, D. Macera, C. Graziani, and M. Ahangarianabhari, "A CMOS charge sensitive amplifier with sub-electron equivalent noise charge," in *Proc. 2014 IEEE Nucl. Sci. Symp.*, Seattle, WA, USA, Nov. 2014, pp. 8–15.

[25] G. Bertuccio and S. Caccia, "Noise minimization of MOSFET input charge amplifiers based on $\Delta\mu$ and $\Delta N 1/f$ models," *IEEE Trans. Nucl. Sci.*, vol. 56, no. 3, pp. 1511–1520, 2009.

[26] R. Hartmann, D. Hauff, S. Krisch, P. Lechner, G. Lutz, R. H. Richter, H. Seitz, L. Struder, G. Bertuccio, L. Fasoli, C. Fiorini, E. Gatti, A. Longoni, E. Pinotti, and M. Sampietro, "Design and test at room temperature of the first Silicon Drift Detector with on-chip electronics," in *IEEE IEDM 1994 Tech. Dig.*, pp. 535–538.